\documentclass[9pt,twocolumn,twoside]{osajnl}
\setboolean{shortarticle}{true}

\definecolor{urlblue}{RGB}{46,46,177}

\journal{ol}

\title{Increased Sensitivity of Higher-Order Laser Beams to Mode Mismatches}

\author[1,2,*]{A. W. Jones}
\author[1,3,4]{A. Freise}
\affil[1]{School of Physics and Astronomy and Institute for Gravitational Wave Astronomy, University of Birmingham, Edgbaston, Birmingham B15 2TT, United Kingdom}
\affil[2]{OzGrav, University of Western Australia, Crawley, Western Australia, Australia}
\affil[3]{Department of Physics and Astronomy, VU Amsterdam, De Boelelaan 1081, 1081, HV, Amsterdam, The Netherlands}
\affil[4]{Nikhef, Science Park 105, 1098, XG Amsterdam, The Netherlands}

\affil[*]{Corresponding author: aaron.jones@ligo.org} 

\begin{abstract}
This manuscript derives explicit factors linking mode-mismatch-induced power losses, in Hermite-Gauss optical modes to the losses of the fundamental spatial mode. Higher order modes are found to be more sensitive to beam parameter mismatches. This is particularly relevant for gravitational-wave detectors, where lasers employing higher-order optical modes have been proposed to mitigate thermal noise and quantum-enhanced detectors are very susceptible to losses. This work should inform mode matching and squeezing requirements for \textit{Advanced+} and \textit{Third Generation} detectors.
\end{abstract}
\begin{document}
\maketitle

\section{Introduction}
Optical higher order modes have a wide range of uses, for example driving micro-machines~\cite{Yao2011,Grier03}, manipulation of cold atoms~\cite{Mestre10} and telecommunications~\cite{Richardson13}. In precision metrology, the performance of current and future gravitational-wave detectors is limited by self-noise of the detectors, which is dominated over a wide frequency band by the quantum noise of the interrogating light field and the thermal noise of the optics.
The introduction of non-classical light (also called \textit{squeezing}) into advanced gravitational-wave detectors~\cite{Tse2019}, leaves 
thermal noise as the fundamentally limiting noise in the detectors' most sensitive frequency range~\cite{AdvancedLIGO15}.

There are proposals to use a spatial Higher-Order Mode (HOM) as the carrier mode in the interferometer to mitigate thermal noise~\cite{Mours06,Vinet07,Chelkowski09} in gravitational-wave detectors. This technique may also be of interest to other thermal-noise limited optical cavities~\cite{Ast2019}. Sorazu et al. studied the use of a Laguerre-Gauss 3,3 (LG33) mode in a 10\,m suspended optical resonator~\cite{Sorazu13} and noted that astigmatism caused the break up of the LG33 mode into component Hermite-Gauss (HG) modes with similar, but not equal round trip Gouy phase, resulting in distorted control signals and a poor power coupling into the resonator.

%
Adaptive astigmatism control could be used to mitigate LG33 break-up~\cite{Vajente13}. Alternatively, Hermite-Gauss modes are naturally astigmatic, and may be more compatible with the long baseline optical resonators used in gravitational-wave-detectors~\cite{Ast2019}. The HG55 has been discussed~\cite{Ast2019} as a possible option. However, it is well known that the transfer of squeezed light into the interferometer is exceptionally sensitive to optical losses~\cite{PhD.Schreiber}. Mode mismatch can be a dominant source of squeezing loss~\cite{Aasi13} and a 98\,\% mode-matching target is achievable with Advanced LIGO+, potentially allowing 8\,dB of squeezing~\cite{Perreca20}.

This paper revisits the subject of higher-order mode to resonator matching, in the context of HG modes, and derives an increased sensitivity to mode mismatch which scales monotonically with mode index. For the HG55, the losses due to waist-size mismatch would be 31 times worse than for the fundamental mode. Results for a waist position mismatch are shown in \textcolor{urlblue}{Supplement 1}.

Our results are derived using a computer algebra system~\cite{sympy} and shown to match numeric integration. A higher-order astigmatic beam passing though the LIGO Output Mode Cleaner is considered as an example of applying these coefficients. These results are consistent with the evidence discussed in~\cite{Sorazu13}, the decreased mode purity and power observed in~\cite{Ast2019} and experimental observations~\cite{PhD.Jones20}.
\section{Analytical Model}
\begin{figure}
	\centering
	\includegraphics[width=\linewidth]{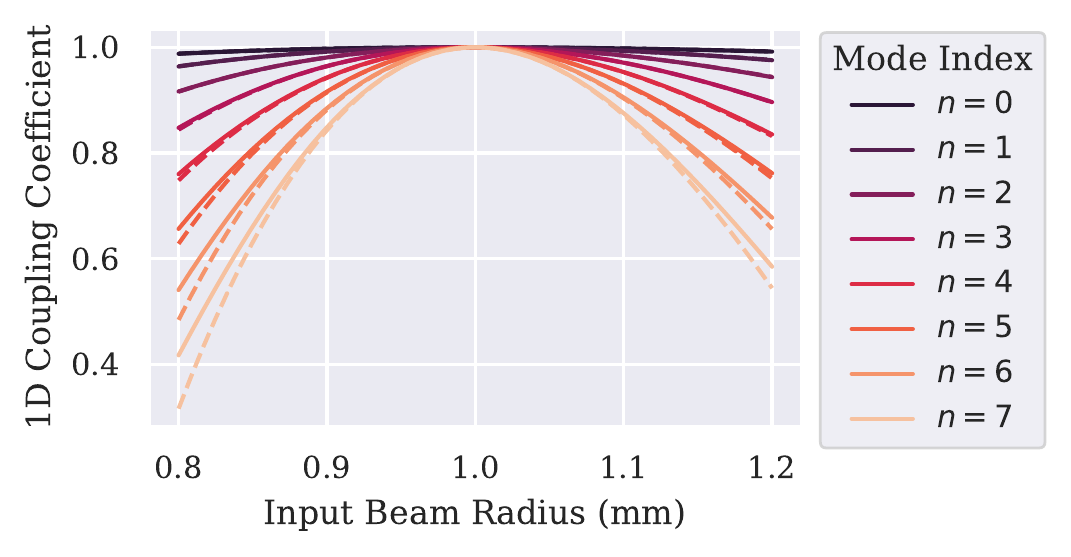}
	\caption{1D mode mismatch parameter, $k_{n,n}$ for a waist size only mismatch between the incoming beam and the 1\,mm resonator waist size. Solid lines show numerical solutions to Equation~\ref{eq:coupling_coeffeciants2} and dotted lines show the approximate analytic solutions in Equation~\ref{eq:approx_solution_mode_mismatch}. See \textcolor{urlblue}{Supplement 1} for an analogous waist-position mismatch.}
	\label{fig:mismatch_analytic_comparison}
\end{figure}
The mode-coupling coefficients were derived in the general case in 1984 by Bayer-Helms~\cite{bayer-helms}. Consider two mode bases, the first with waist $w_0$ at $z_0$ (typically the mode basis of the incoming light) and the second with waist $\overline{w}_0$ at $\overline{z}_0$ (typically the resonator mode basis). Then, the amplitude coupling of a mode with indices $n,m$ in the first basis, to mode $\overline{n},\overline{m}$ in the second basis is described by, $k_{{n},{m},\overline{n},\overline{m}}$, which is in general complex. In this work, all parameters with an overline correspond to the resonator basis.

For Hermite-Gauss modes, this coupling coefficient is separable~\cite{bayer-helms},
\begin{align}
k_{{n},{m},\overline{n},\overline{m}} = k_{{n},\overline{n}}k_{{m},\overline{m}}.
\end{align}
If the beam axis is aligned, then the 1D coupling coefficients are reduced to~\cite{bayer-helms},
\begin{align}
k_{n,\overline{n}} =
\int_{-\infty}^\infty
{u}_{{n}}\left({x'},{z}\right)
\overline{u}_{\overline{n}}^* \left({x'},{z}\right)
\mathrm{d}\!x'.
\label{eq:coupling_coeffeciants2}
\end{align}
Considering only, $w_0 \neq \overline{w}_0, z_0 = \overline{z}_0$ (See \textcolor{urlblue}{Supplement 1} for $w_0 = \overline{w}_0, z_0 \neq \overline{z}_0$), then evaluating both beams at the waist $z=z_0$, the beam size is $w(z)=w_0$ and the radius of curvature is $R_C = \infty$. Additionally, the Gouy phase of the resonator mode at the waist is zero $\overline{\Psi}(z_0)=0$. Then, by rescaling $x=x'/\overline{w}_0$, the spatial properties of the resonator eigenmodes are~\cite{Bond2017},
\begin{align}
\overline{u}_{\overline{n}}(x,z) = 
\left(\frac{2}{\pi}\right)^\frac{1}{4}\!
\sqrt{\frac{1}{2^{\overline{n}} \overline{n}!}}
H_{\overline{n}}\left(\sqrt{2}x\right)e^{-x^2}.
\label{eq:resonator_field}
\end{align}
Defining the fractional waist size mismatch, $w \equiv w_0/\overline{w_0}$, the distribution of the incoming light is,
\begin{align}
{u}_n(x,z) = \left(\frac{2}{\pi}\right)^\frac{1}{4}\sqrt{\frac{1}{2^n n! w}}
\exp\left(\!\frac{i(2{n}+1)\Psi(z)}{2}\right)\nonumber & \\
H_n\left( \frac{\sqrt{2}x}{w}\right)\exp\left(\frac{-x^2}{w^2}\right),&
\label{eq:incoming_field}
\end{align}
where $\Psi$ is free to describe some accumulated Gouy phase. After substitution of Equations~\ref{eq:incoming_field} \& \ref{eq:resonator_field}, Equation~\ref{eq:coupling_coeffeciants2} is difficult to solve. However, the \texttt{integrate} function from SymPy~\cite{sympy} v1.3, can solve this for a specific $n$, which may then be expanded with the \texttt{series} method. For the first 10 orders, the coupling constant between the same mode in each basis ($n=\overline{n}$) is,
\begin{align}
k_{n,n} \approx& \exp\left(\frac{i(2n+1)\Psi(z)}{2}\right)\nonumber\\&
\left(1 - \frac{C_{n}}{4}\left((w-1)^2 - (w-1)^3\right) + \mathcal{O}\left((w-1)^4\right)\right),
\label{eq:approx_solution_mode_mismatch}
\end{align}
where,
\begin{align}
C_0 = 1, \quad C_1 = 3, \quad C_2 = 7, \quad C_3 &= 13,\quad C_4 = 21,\quad
C_5 = 31,\nonumber \\ 
C_6 = 43, \quad C_7 = 57, \quad C_8 = 73, &\quad C_9 = 91,
 \quad C_{10} = 111, 
\end{align}
and \textcolor{urlblue}{Code File 1} (Ref. \cite{supl2}) can be used to compute additional values of $C_n$.
Figure~\ref{fig:mismatch_analytic_comparison} shows a numerical solution to Equation~\ref{eq:coupling_coeffeciants2} using PyKat~\cite{PyKat_Paper} against Equation~\ref{eq:approx_solution_mode_mismatch} expanded to order $(w-1)^3$. For a waist size mismatch less than 5\,\% there is good agreement between the analytic solution and the numerical ones. 

When considering a resonator, the power coupling efficiency, $k_{n,\overline{n},m,\overline{m}}k^*_{n,\overline{n},m,\overline{m}}$, is considered. Defining the horizontal losses to be,
\begin{align}
W_x \equiv \frac{(w-1)^2 - (w-1)^3}{4} \approx 1-\left|k_{0,0}\right|
\end{align}
and likewise for the vertical losses, $W_y$, the full 2D coupling coefficient is,
\begin{align}
k_{n,{n},m,{m}} 
\approx e^{i({n}+{m}+1)\Psi(z)}(1-C_nW_x - C_mW_y + C_nC_mW_xW_y).
\end{align} 
For an almost matched beam in $x$ and $y$, the last term may be safely ignored. The power coupling coefficient is then,
\begin{align}
k_{n,{n},m,{m}}k^*_{n,n,m,{m}} \approx 1-2C_nW_x - 2C_mW_y,
\end{align}
where terms of order $W_x^2$, $W_y^2$ and $W_xW_y$ have been neglected. This result conclusively shows that, higher-order modes are more susceptible to mode mismatching losses when coupling into cavities. 

\section{Example - Power Throughput of the Advanced LIGO Output Mode Cleaner}
\begin{figure}
\centering
\includegraphics[width=\linewidth]{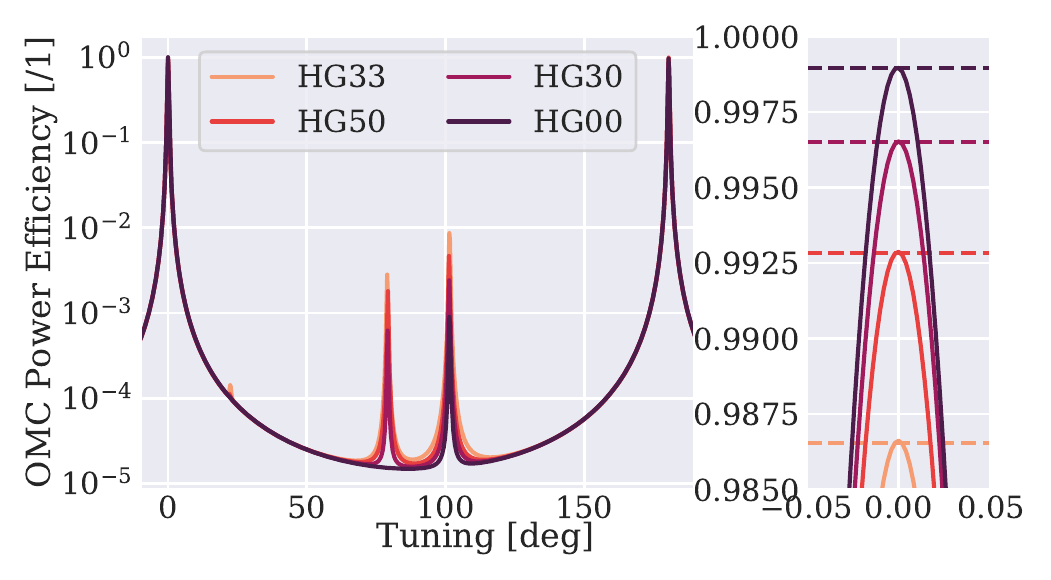}
\caption{Power transmitted by the Advanced LIGO OMC for an astigmatic input beam. On both plots, the $x$ axis shows tuning from expected resonance position, the $y$ axis shows the transmitted power. The right hand plot shows a zoom of the peak resonance on a linear scale, dashed lines show efficiency determined with Equation~\ref{eq:knnmm_effect}.}
\label{fig:OMC_modemis}
\end{figure}
\begin{table*}
\centering
\begin{tabular}{c | c c c c c}
Input Mode & Analytic ($x$) & Analytic ($y$) & Analytic (Total) & Simulation & Difference
\\ \hline
HG00 & 204.0\,ppm & 832.6\,ppm & 1036.7\,ppm & 1036.5\,ppm & 0.2\,ppm\\
HG30 & 2652.5\,ppm & 832.6\,ppm & 3485.1\,ppm & 3472.8\,ppm & 12.3\,ppm\\
HG50 & 6325.1\,ppm & 832.6\,ppm & 7157.8\,ppm & 7131.6\,ppm & 26.1\,ppm\\
HG33 & 2652.5\,ppm & 10824.1\,ppm & 13476.6\,ppm & 13389.8\,ppm & 86.8\,ppm
\end{tabular}
\caption{Mode mismatch induced power losses through the OMC for an astigmatic input beam. The analytic response is determined from Equation~\ref{eq:knnmm_effect} and the simulated response is determined from the \texttt{Finesse} cavity scan in Figure~\ref{fig:OMC_modemis}.} 
\label{tab:OMC_modemis}
\end{table*}
Advanced LIGO operates with a high degree of mode matching to ensure power couples efficiently between the resonators; however, some degree of mismatch is always present.

Within the core interferometer an increased sensitivity to mode mismatch will likely cause a reduced interferometric visibility. In addition, since the core interferometer is dual recycled and has focusing elements within the recycling cavities, an increased sensitivity to mode mismatch may lead to challenges in defining an operating point for the resonators~\cite{phd.bond2014}. 

In the case of the Input and Output Mode Cleaners (IMC and OMC), modes which are not resonant are reflected and dumped. Therefore, the effect of the mode mismatch is a reduced power transmission through the resonator. In the case of the IMC, small mismatches can be compensated for by increasing laser power. In the case of the OMC the mode mismatch directly causes a loss of signal and loss for squeezed light injection.

A \texttt{Finesse} model~\cite{Freise04, Finesse} of the Advanced LIGO OMC was produced and the transmission efficiency was studied for a range of input modes, results are shown in Figure~\ref{fig:OMC_modemis}. The input power was chosen such that a mode-matched beam produced 1\,W of power on transmission, when the resonator was tuned and was constant for all simulations. This power scaling means that the power on transmission is equal to the OMC power coupling efficiency. The input beam was astigmatic with $w_{0x} = 0.98\overline{w}_{0x}$ and $w_{0y} = 0.96\overline{w}_{0y}$. This astigmatism was chosen to highlight the differing losses for modes with $m\neq n$.The tuning range was measured from the expected resonance position. Simulation modes $n',m'$, up to $n'+m' \leq n+m+4$, for input mode $n,m$ were enabled. 

The parameter $2W_x$ was determined by running an additional simulation with TEM00 input and $w_{0y} = \overline{w}_{0y}$ and $w_{0x} = 0.98\overline{w}_{0x}$, then
\begin{align}
2W_x = 1 - \frac{P_{Tx}}{P_{T}},
\end{align}
where $P_{Tx}$ is the power measured on transmission and $P_{T}$ is the transmitted power for no mismatch. In this work, the input power scaling meant $P_{T}=1$. The parameter $2W_y$ was obtained similarly. The analytically determined OMC power coupling efficiency for mode HGnm is then,
\begin{align}
\left|k_{n,{n},m,{m}}\right|^2 = 1 - C_n\left(1-\frac{P_{Tx}}{P_{T}}\right) - C_m\left(1-\frac{P_{Ty}}{P_{T}}\right),
\label{eq:knnmm_effect}
\end{align}
which is shown by the dotted lines in Figure~\ref{fig:OMC_modemis}. This general method also works as an experimental procedure and can be used to estimate losses in switching to a higher-order mode.

$\left|k_{n,{n},m,{m}}\right|^2$ was also obtained directly from the simulation by measuring the peak transmitted power, a comparison is shown in Table~\ref{tab:OMC_modemis}. As an example, when the $n$ index is increased from 0 to 3, the $x$ related power losses increase by 13 times. When the $m$ index is increased as well, both $x$ and $y$ power losses increase, so the total mode mismatch induced power loss increases by 13 times.

Mode-mismatch-induced power losses in the OMC correspond directly to a loss of signal and increased quantum noise. Changing to an equivalently stable higher-order spatial mode will reduce thermal noise, however, unless the higher-order mode matching is improved compared to the TEM00 mode matching, the mode mismatching induced signal degradation will be 13 times worse for a HG33 and 31 times worse for a HG55 carrier mode.
\section*{Funding} Lorium Ipsum

\section*{Acknowledgments} The authors jointly thank Dr Chris Collins and Dr Conor Mow-Lowry for helpful discussions.

\section*{Disclosures}
The authors declare no conflicts of interest.


\noindent See \textcolor{urlblue}{Supplement 1} for supporting content.
\bibliography{extracted}





\bibliographyfullrefs{extracted}
\end{document}


\maketitle

\section{Introduction}

The focus of the main text is the determination of the increased sensitivity of a higher-order laser beam to mode mismatches. The main text uses a waist size mismatch as an example and derives coefficients describing the increase in sensitivity to mode mismatch. However, similar coefficients can also be derived for waist position mismatches. Beam misalignment has not been considered as gravitational wave detectors use resonant wavefront sensing to maintain the alignment \cite{Morrison1994}.

\section{Waist Position Mismatch Formulation}
First, we define a normalized waist position mismatch parameter (which is equal to $K_0$ in \cite{bayer-helms}),
\begin{align}
f = \frac{z_0 - \overline{z_0}}{z_R}.
\end{align}
Without loss of generality, the resonator waist position is defined to be ($\overline{z}_0 = 0$) and the resonator Gouy phase is zero at the waist. The waist sizes are assumed to be the same, therefore $z_R=\overline{z}_R$. Then, the incoming beam radius and radius of curvature at $\overline{z_0}$ may then be rewritten in terms of the mismatch parameter $f$,
\begin{align}
R_C(f) &= z_R \left(f + 1/f\right),\\
w^2(f) &= \frac{2z_R}{k}\left(1+f^2\right).
\end{align}
Substitution of these parameters into,
\begin{align}
k_{n,\overline{n}} =
\int_{-\infty}^\infty
{u}_{{n}}\left({x},{z}\right)
\overline{u}_{\overline{n}}^* \left({x},{z}\right)
\mathrm{d}\!x.
\label{eq:coupling_coeffeciants2}
\end{align}
then yields the coupling coefficients. We found that a series expansion needed to be taken prior to integrating Equation \ref{eq:coupling_coeffeciants2}, this series expansion was taken to order $f^7$. Solving using a symbolic library then yields,
\begin{align}
k_{n,n} \approx \exp\left(\frac{i(2n+1)\Psi(z)}{2}\right)
\left(1+\frac{iC_n^\text{I}f}{4} - \frac{C_n^\text{II}f^2}{32}\right) + \mathcal{O}(f^3),
\label{eq:approx_solution_mode_mismatch}
\end{align}
where, for the first 10 orders, the linear coefficients are just $C_n^\text{I} = 2\text{n}+1$ and the quadratic coefficients are,
\begin{align}
C_0^\text{II} = 3, \quad C_1^\text{II} = 15, \quad C_2^\text{II} = 39,& \quad C_3^\text{II} = 75,\quad C_4^\text{II} = 123,
\quad C_5^\text{II} = 183,\nonumber \\ 
\quad C_6^\text{II} = 255, \quad C_7^\text{II} = 339, \quad C_8^\text{II} &= 435, \quad C_9^\text{I} = 543, \quad C_{10}^\text{I} = 663.
\end{align}
\section{Numerical Comparison}
\begin{figure}
\centering
\includegraphics{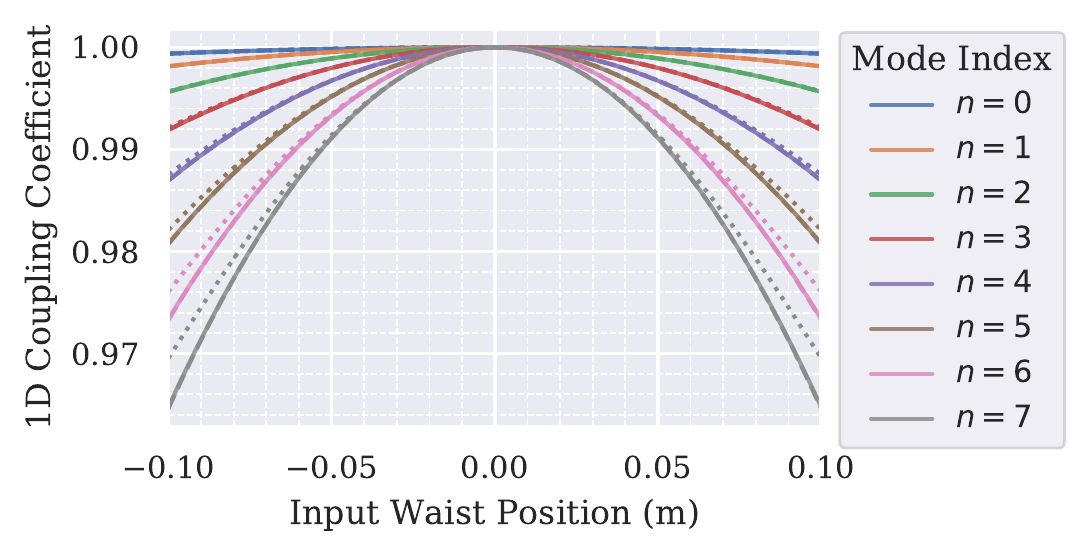}
\caption{1D mode mismatch parameter, $k_{n,\overline{n}}$ for a waist position only mismatch between the incoming beam. Both beams have $z_R=1$\,m. Solid lines show numerical solutions to Equation~\ref{eq:coupling_coeffeciants2} and dotted lines show the approximate analytic solutions in Equation~\ref{eq:approx_solution_mode_mismatch}. This result is generalizable for other waist radii by normalizing the mismatch by the resonator Rayleigh range.}
\label{fig:waist_pos}
\end{figure}
Figure~\ref{fig:waist_pos} shows a numerical solution to Equation~\ref{eq:coupling_coeffeciants2} using PyKat\cite{PyKat_Paper}. For a normalized waist position mismatch less than 5\,\% there is good agreement between the analytics and the numerics for all modes. As the mode index increases, more terms are required to model a given position mismatch.

\section{Conclusions}
Both waist position and waist size mismatches result in increasing mode mismatching losses as the mode order is increased.

\bibliography{extracted}
